\UseRawInputEncoding

\documentclass[%
 reprint,
 amsmath,amssymb,
prl,
]{revtex4-2}

\usepackage{lipsum}
\usepackage{graphicx}
\usepackage{bm}
\usepackage{color}
\usepackage{times}
\usepackage{amsmath}
\usepackage{float}
\usepackage{hyperref}
\usepackage{ulem} 
\usepackage[dvipsnames]{xcolor}

\newcommand{\beginsupplement}{%
	\setcounter{table}{0}
	\renewcommand{\thetable}{S\arabic{table}}%
	\setcounter{figure}{0}
	\renewcommand{\thefigure}{S\arabic{figure}}%
 	\setcounter{equation}{0}
	\renewcommand{\theequation}{S\arabic{equation}}%
	\setcounter{page}{0}
	\renewcommand{\thepage}{S\arabic{page}}%
	
}

\begin{document}


\title{Harnessing curvature for helical wave generation in spiral-based metamaterial structures}

\author{Mohamed Roshdy}
\author{Osama R. Bilal  }
\email{osama.bilal@uconn.edu}
 \affiliation{School of Mechanical, Aerospace, and Manufacturing Engineering, University of Connecticut, Storrs , CT , 06269, USA.}
 
\date{\today}

\begin{abstract}
{Linearly polarized elastic waves propagating in a linear path have been extensively explored in numerous applications from biomedical imaging to structural health monitoring. However, elastic waves propagating in a circularly polarized helical path are less explored because of the challenges in their generation and control.  In this paper, we harness conventional actuation methods combined with sheets decorated by Archimedean spirals to generate elastic helical waves. We show that our metamaterials can support the propagation of such waves along a curved path without backscattering in a topologically protected manner. Moreover, we also show the creation and propagation of helical waves in our metamaterials without the need to wave guiding or domain interfaces, all with a single mode excitation source. We establish our methodology for flat plates and show the wave evolution as the metamaterials transition from slightly curved plates to fully curved cylinders. We observe the preservation of topologically protected edge states and helical wave beaming at different frequencies without the need to domain interferences. Furthermore, we leverage the mode to tune the helicity of the propagating waves. Our methodology can open new avenues for the generation and control of elastic helical waves using single source actuation that can be used in numerous applications.}

\end{abstract}

\maketitle

The propagation of elastic waves in solid media can involve the coupling of both volumetric and shear mechanical deformations of the particles within the solid. In a conventional isotropic solid, the wave propagation speeds are vastly different for longitudinal versus shear polarizations. This poses a challenge to the generation and control of polarized elastic waves in solids \cite{long2018intrinsic,yuan2021observation,lee2024control, lee2025mechanical,yu2018elastic,lee2024perfect}, in contrast to electric or magnetic waves propagating in free space or acoustic waves propagating in a fluid medium \cite{shi2019observation, long2020realization}. In addition, for an elastic wave to travel in a circularly polarized helical path, both the shear and longitudinal wave components must be locked with a 90$^\circ$ phase difference. Due to the complexity in generating and supporting the propagation of helical waves in solids, i.e., waves propagating in circular inclined wavefronts, the utility of these waves is far less explored in both the literature and practice. Moreover, in most cases, helical waves are considered in Cartesian coordinates with little to no regard to curvature. Curvature, a prevalent trait in both natural and engineered systems, introduces more challenges in the analysis, modeling, and control of such waves \cite{conoir1993relation,marston1997spatial,thothadri1998helical,haumesser2001behavior,blonigen2002leaky,perton2006bulk,leonard2003guided, tyutekin2004helical, balvantin2013study, baltazar2015structural, oukhai2024behaviour, sadowski2014modelling, senga2017forced, manconi2018wave,treyssede2008elastic, treyssede2007numerical, treyssede2010investigation, treyssede2011mode}.

\begin{figure}
\centering
\includegraphics[width = \columnwidth]{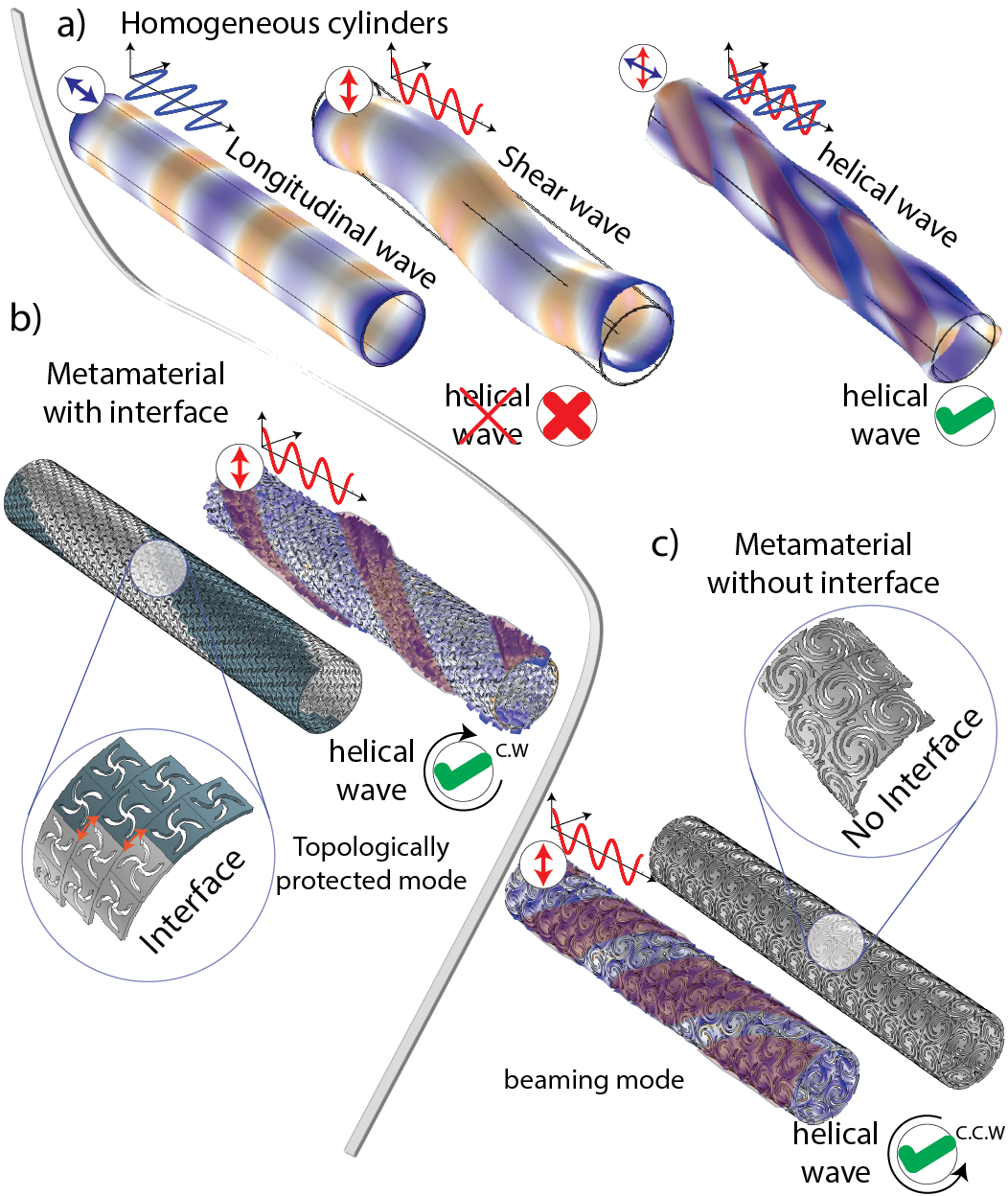}
\caption{\label{fig:Concept}\textbf{ Helical wave generation in a cylinder}. a)  Longitudinal and shear (transverse) waves in a homogeneous cylinder and helical wave excitation with complex phase-locking in a homogeneous cylinder. Simple single mode excitation of helical wave with  b) topologically protected interface along a metamaterial cylinder and  c) a uniform (no interface) metamaterial cylinder with an intrinsic beaming mode.
}

\end{figure}

Here, we propose the utilization of metamaterials to generate and support the propagation of helical waves. Metamaterials are artificial, periodic structures designed to exhibit unconventional properties, as a result of their geometric design rather than their chemical composition. These unconventional properties include negative effective Poisson's ratio, negative effective mass, or negative effective stiffness \cite{deshpande2001effective, greaves2011poisson, huang2009negative, zhou2012elastic}. Metamaterials have been utilized to guided acoustic helical wavefronts in fluids \cite{zhu2016implementation,esfahlani2017generation,fan2019tunable}, transfer angular momentum to a fluid in contact with a solid \cite{chaplain2022elastic, chaplain2022elastic2}, or spin-orbit coupling in helically curved rods \cite{yang2024chirality}.  Despite these attempts, the generation of helical waves in curved structures remains challenging due to the complexity of the required designs and excitation methods \cite{lee2024perfect}. Metamaterials have been utilized in numerous applications, such as energy harvesting \cite{lv2013vibration,qi2016acoustic,jo2020elastic}, actuation and sensing \cite{gallego1989piezoelectric,vasile1979excitation}, and imaging \cite{li2009experimental} with the majority of the literature focusing on linear longitudinal or transverse elastic guided waves. The simplicity of our proposal can be illustrated by considering a homogeneous cylinder excited with a single-polarization actuator [Fig. \ref{fig:Concept} a]. The cylinder can oscillate either axially, in the direction of the wave propagation (i.e., longitudinal waves), or transversely, perpendicular to the direction of the wave propagation (i.e., shear waves). To introduce helicty, a precise 90$^\circ$ out-of-phase excitation with longitudinal and transverse excitation is usually utilized, resulting in helical waves. Instead, our approach is to decorate the cylinder with a topologically protected spiraling interface enabling the generation of backscattering-immune helical waves with a controlled number of helical turns [Fig. \ref{fig:Concept} b]. In addition, in case spatial uniformity is necessary, we show an alternate method to generate helical waves in a uniform meta-cylinder without the need for a waveguide through beaming [Fig. \ref{fig:Concept} c]. Through our approach, both the number of turns within the wave helix and its rotation direction (clockwise or counterclockwise) can be tuned.

\section{RESULTS AND DISCUSSION}
In order to elucidate our methodology, we start by sculpting the wave dispersion relation through rational design. Metamaterial dispersion band diagrams can be used to predict and engineer unusual dynamical characteristics, such as steering elastic waves to propagate along a specific path rather than dispersing in all directions (i.e., wave beaming) \cite{langley1996response}, or to propagate only along the edges of the structure while remaining robust against defects, disorder, or perturbations (i.e., topologically protected) \cite{kane2005quantum}. Within these dispersion bands, frequency gaps can evolve as a function of the metamaterial's basic building block (i.e., unit cell). Band gaps for elastic waves can open due to different physics phenomena, including Bragg scattering, local resonance, and inertia amplification. Each of these band gap opening mechanisms usually requires a specific design strategy. However, there exists a unified approach to utilize all these band gaps based on introducing Archimedean spiral cuts within a flat plate \cite{foehr2018spiral}. By changing the spiral parameters, different patterns emerge, leading to the opening of various types of band gaps. This approach remains limited to flat spiral-based metamaterials and without experimental realization \cite{foehr2018spiral}. Here, we extend the design framework to curved structures and experimentally validate the dynamic behavior of the resulting designs (both flat and curved). The spiral parameters are outer radius $R$, inner radius $r$, cutting width $W$, orientation angle $\alpha$, and number of turns $n$. We consider four different spiraling metamaterials designs that exhibit different elastic band gap opening mechanisms. First we consider Bragg scattering, where waves are being prohibited from propagation due to the destructive interference at wavelengths of the same order of the unit cell spacing. Bragg scattering band gaps exist in Designs 1 and 2 (See supporting information [Fig. \ref{fig:S1} a-b]). The second mechanism for opening the band gap is local resonance, where the band gaps originate from the localization of energy at certain geometric features. These features resonate at sub-wavelength frequencies. The local resonance band gap exists in Design 3 (See supporting information [Fig. \ref{fig:S1} c]). The third band gap opening mechanism is inertia amplification, where the amplification of the resonator effective inertia opens the band gap. The inertia amplification band gap exists in Design 4 (See supporting information [Fig. \ref{fig:S1} d]).

\subsection{ TOPOLOGICALLY PROTECTED INTERFACES}
\begin{figure*}
\centering
\includegraphics[width=0.98\textwidth]{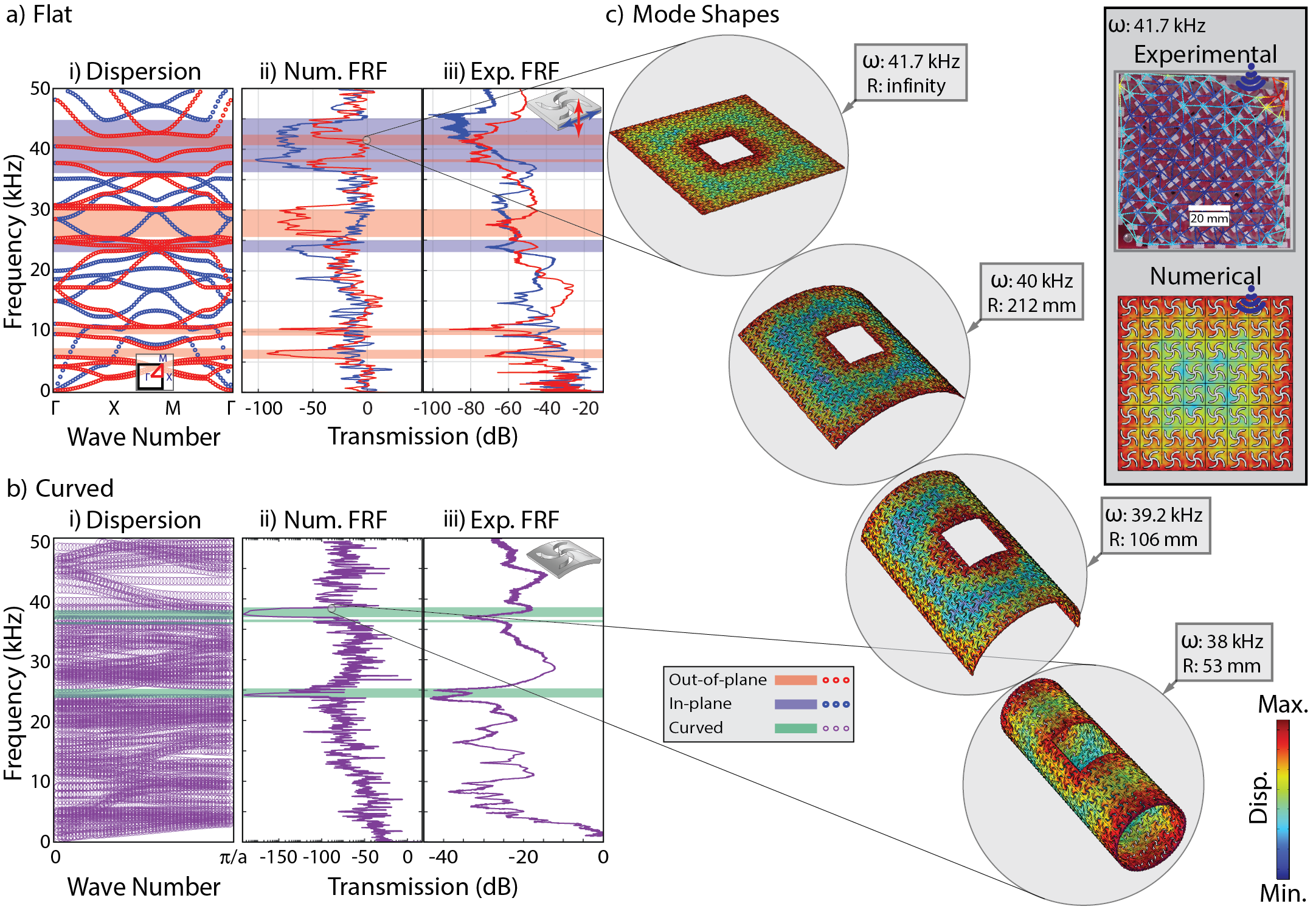}
\caption{\label{fig:Topological mode}\textbf{Topological mode evolution (Design 1).} Dispersion curves, numerical, and experimental frequency response functions (FRF) for Design 1 in  a) Flat and b) curved settings. Inset in panel a(i) shows the irreducible Brillouin zone (IBZ) $\Gamma-X-M-\Gamma$. In the flat dispersion, there are two in-plane band gaps at $\omega_c$ of 24 and 40.66 kHz with $BG \%$ of 8 and 21.8 \%, respectively, and five out-of-plane band gaps at $\omega_c$ of 6.5, 10, 27.9, 38.1, and 41.5 kHz with $BG \%$ of 22, 11.2, 16.1, 1.1, and 4.1 \%, respectively. In the cured dispersion, there are three band gaps at $\omega_c$ of 24.5, 36.4, and 37.8 kHz with $BG \%$ of 6.1, 1.1, 4.2 \%, respectively. c) Evolution of the topologically protected edge modes for a 16x16 flat plate  with 4x4 unit cells removed from its center as it turns into a full cylinder with radius of curvature reaching 53 mm. Inset in top-left corner shows numerical and experimental mode shapes of the topologically protected mode at 41.7 kHz for a flat plat of 8x8 unit cells.}
\end{figure*}

To understand the dynamics of the considered designs, we calculate the dispersion diagrams for each design as a single unit cell repeating infinitely in space. We start with Design 1 in Cartesian coordinates (x,y,z) as a flat unit cell. To account for the infinite periodicity of the unit cell, we consider Bloch's periodic boundary conditions in both x- and y-directions, where the solution of the wave equation is assumed to be in the form $u_{car}(\textbf{x},\kappa;t) = \bar{u}_{car}(\textbf{x},\kappa) e^{i (\kappa \textbf{x} - \omega t)}$ \cite{bloch1929quantenmechanik}, where $\bar{u}_{car}$  is the displacement Bloch function, $\textbf{x}$ is the position vector in Cartesian coordinates, $t$ is the time, and $i$ is the complex number $\sqrt{-1}$. We use two metrics to quantify the resulting band gaps; the band gap central frequency $\omega_c = (\omega_{max} + \omega_{min})/2$, and the band gap percentage ($BG \% = \Delta \omega/\omega_c$), where $\omega_{max}$ is the band gap upper frequency limit, $\omega_{min}$ is the band gap lower frequency limit, and $\Delta \omega$ is the band gap width. We calculate the dispersion curves for both in-plane and out-of-plane modes and observe two in-plane band gaps and five out-of-plane band gaps [Fig. \ref{fig:Topological mode} a (i)]. Our main focus here is on the band gap centered at 41.7 kHz for its topologically protected properties. 

\begin{figure*}
\centering
\includegraphics[width= 0.95\textwidth]{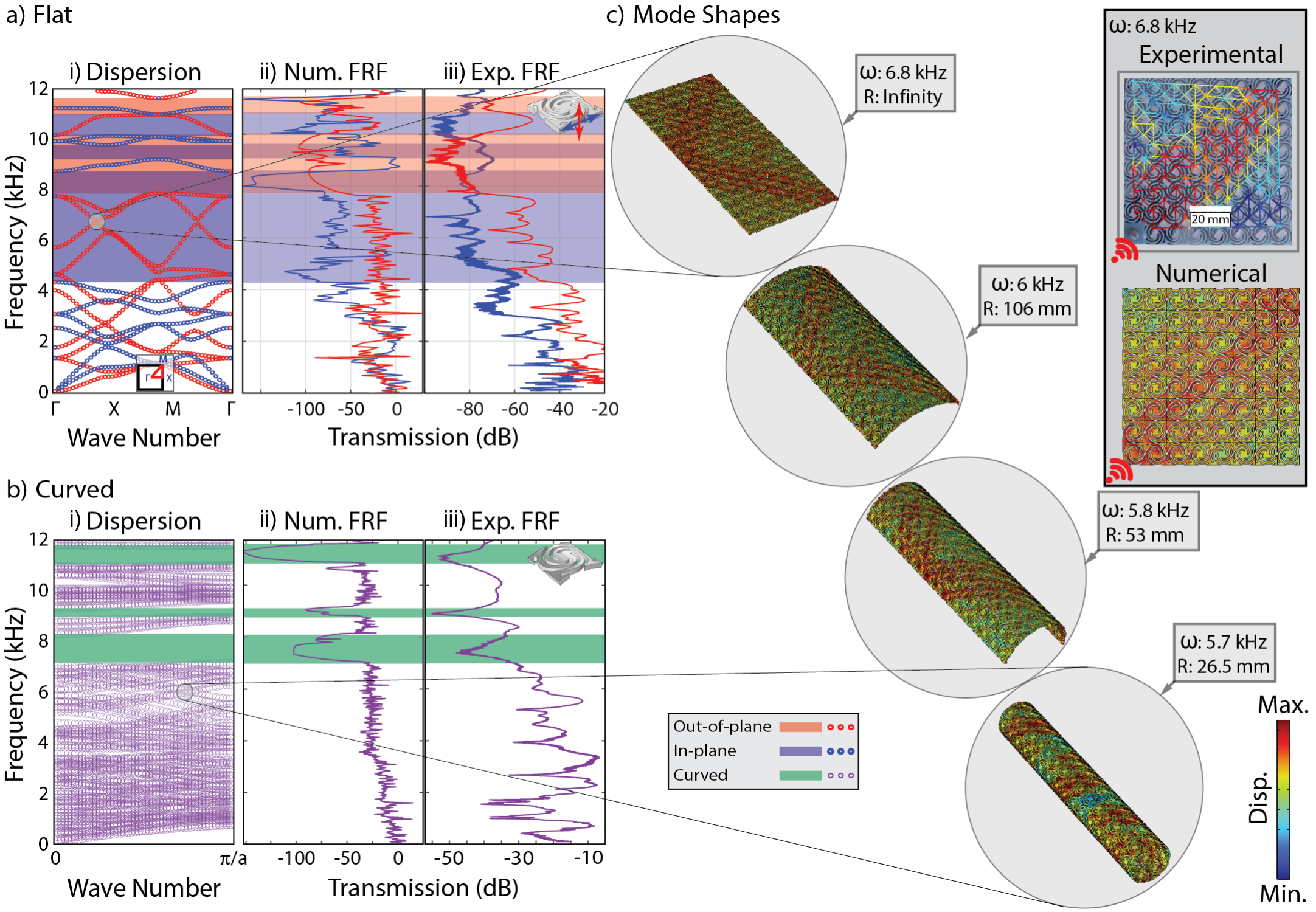}
\caption{\label{fig:Beaming mode}\textbf{Beaming mode evolution (Design 2).} Dispersion diagram, numerical, and experimental frequency response functions (FRF) For Design 2 in a) Flat  and b) curved settings. In the flat dispersion, there are three in-plane band gaps at $\omega_c$ of 6.44, 9.35, and 10.41 kHz with $BG \%$ of 67, 6, and 8.2 \%, respectively and two in-plane band gaps at $\omega_c$ of 8.89 and 11.14 kHz with $BG \%$ of 26 and 6 \%, respectively.  In the cured dispersion, there are three band gaps at $\omega_c$ of 7.55, 8.95, and 11.2 kHz with $BG \%$ of 15, 4, and 7 \%, respectively. c) Evolution of the beaming modes for an 8x16 flat plate as it turns into a full cylinder with radius of curvature reaching 26.5 mm. Inset in top-right corner shows numerical and experimental mode shapes of the beaming mode at 6.8 kHz for a flat plate of 8x8 unit cells.}
\end{figure*} 

To demonstrate the effect of curvature on the dynamics of metamaterials, we consider the same unit cell of Design 1 but curved over an angle of $45^{o}$ in the cylindrical coordinates (r,$\theta$,z). This value of curvature results in eight unit cells in the circumferential directions $N = 8$. We calculate the dispersion diagrams of the curved unit cells considering infinite periodicity in the axial direction and finite periodicity in the circumferential direction (i.e., cyclic periodicity). We need to change the wave solution to the form $u_{cyl}(\bar{\textbf{R}},\kappa;t) = \bar{u}_{cyl}(\bar{\textbf{R}},\kappa) e^{i(\kappa_z n_z a -\omega t)} e^{i m\theta}$ \cite{roshdy2025static}, where $\bar{\textbf{R}}$ is the position vector in the cylindrical coordinates, $\kappa_z$ is the wave number in the axial direction, $n_z$ in the unit cell index in the axial direction, $\theta$ is the cylindrical coordinate angle, and $m$ is the circumferential mode number (i.e., Azimuthal mode number). The circumferential mode number $m$ takes the value from 0 to $N/2 \:$ with an increment of 1. We calculate the dispersion curves in the axial direction over the wave vector $\kappa_z$ = $0$ to $\pi/a$, for all the circumferential mode numbers (i.e., $m = 0\:to\:4$). It is important to note that there are no more in-plane or out-of-plane modes, but rather coupled modes. Therefore, we focus on the aggregate band gaps, where all circumferential modes have common band gaps. With both shear and longitudinal modes coupled, we observe three band gaps [Fig. \ref{fig:Topological mode} b (i)]. More importantly, we observe a clear change in the band gaps' size and location in the frequency spectrum from the flat to the curved dispersion diagrams.

To validate our infinite dispersion models, we construct finite structures out of the considered unit cell of Design 1, both in flat and curved configuration. We excite the finite structures at one end with a chirp signal and measure the transmission of the wave at the other end of the structure. The resulting frequency response functions (FRFs) show a very good agreement with the infinite model prediction, particularly at the shaded band gap regions coinciding with the low displacement frequency regions (attenuation regions) for both the flat [Fig. \ref{fig:Topological mode} a (ii,iii)] and curved case [Fig. \ref{fig:Topological mode} b (ii,iii)]. Design 1 has a band gap with a degenerate Dirac cone (at 41.7 kHz), which enables us to obtain a topologically protected edge mode, where elastic wave propagates along the edges of the structure while being protected from defects and imperfections [Fig.\ref{fig:S4} a]. The edge mode appears in both numerical and experimental FRFs as a high displacement peak at 41.7 kHz [Fig. \ref{fig:Topological mode} a (ii,iii)]. The numerical and experimental mode shapes of 8x8 unit cells plate show a topological edge mode propagating at the edge of the plate [Fig. \ref{fig:Topological mode} c top right corner]. We further construct a 16x16 flat plate of Design 1 and add a defect in the middle of the plate by removing 4x4 unit cells. The mode shape at 41.7 kHz shows a topological edge state mode with high displacement amplitude only at the edges of the structure. As we curve the plate, from partially curved plates, to a fully curved cylinder, we observe a preservation of the topological edge mode for all structures but at lower frequencies [Fig. \ref{fig:Topological mode} c]. The frequency of the edge mode evolves gradually from 41.7 kHz for the flat design to 38 kHz, which agrees well with a high displacement peak inside the curved unit cell band gap. The mode shapes confirm that a topologically protected edge state mode in a flat metamaterial plate is preserved with the plate curving but at different frequencies.

\subsection{SPIRAL DESIGN WITH BEAMING MODE}
In addition to their ability to guide waves without backscattering in topologically protected waveguides, the intrinsic chiral nature of spiraling metamaterials can enable wave beaming as a function of the excitation frequency. To harness wave beaming, we vary the parameters of the spiraling unit cell systematically to reach Design 2.  We perform a similar dynamic analysis to the previous design by calculating the infinite dispersion diagrams for a flat and a curved unit cell of the same Design 2. In the flat case, we observe three in-plane and two out-of-plane band gaps [Fig. \ref{fig:Beaming mode} a (i)]. For the curved unit cell, we observe three band gaps [Fig. \ref{fig:Beaming mode} b (i)]. Notably, there is a clear change in width and location of the band gaps from flat to curved, confirming the importance of the geometric effect of curvature in metamaterial \cite{roshdy2025static}. We further validate our prediction by testing finite structures made out of Design 2, both in flat [Fig. \ref{fig:Beaming mode} a (ii,iii)] and curved settings [Fig. \ref{fig:Beaming mode} b (ii,iii)]. Both numerical and experimental FRFs show a very good agreement with the dispersion diagram, with the shaded band gaps matching well with the low displacement regions in the FRFs.

The dispersion diagram of the flat Design 2 shows a beaming mode at 6.8 kHz, where only a single out-of-plane mode exists in the reciprocal space $\Gamma$ to $X$ (See Supp. Info. [Fig.\ref{fig:S5} a]). Both the numerical simulations and experimental measurements of an 8x8 finite structure show diagonal wave beaming from the excitation point to the diagonally opposite corner [Fig. \ref{fig:Beaming mode} c - top right corner]. We further consider a finite structure composed of 8x16 unit cells (aspect ratio 1:2). When excited at the right corner with the same 6.8 kHz frequency, the wave beam follow the diagonal of the lower half of the structure, after which it reflects at the midpoint of the left side and propagates through the opposite diagonal reaching the opposite right corner. As the flat plate curves into a full cylinder, the beaming mode persists but shifts to lower frequencies. Intriguingly, the beaming mode generates a helical wave around the surface of the cylinder without any artificial waveguides [Fig. \ref{fig:Beaming mode} c].

\subsection{HELICAL WAVE TUNABILITY}
To demonstrate the versatility of helical wave generation in our framework, we utilize the preserved topologically protected edge mode in curved spiral Design 1 and the persisting beaming mode in curved spiral Design 2 to generated helical waves that propagate on the surface of the metamaterial cylinders. In addition, we tune the number of helical turns per certain cylinder length (i.e., helicity). 

\begin{figure}
\centering
\includegraphics[width= \columnwidth]{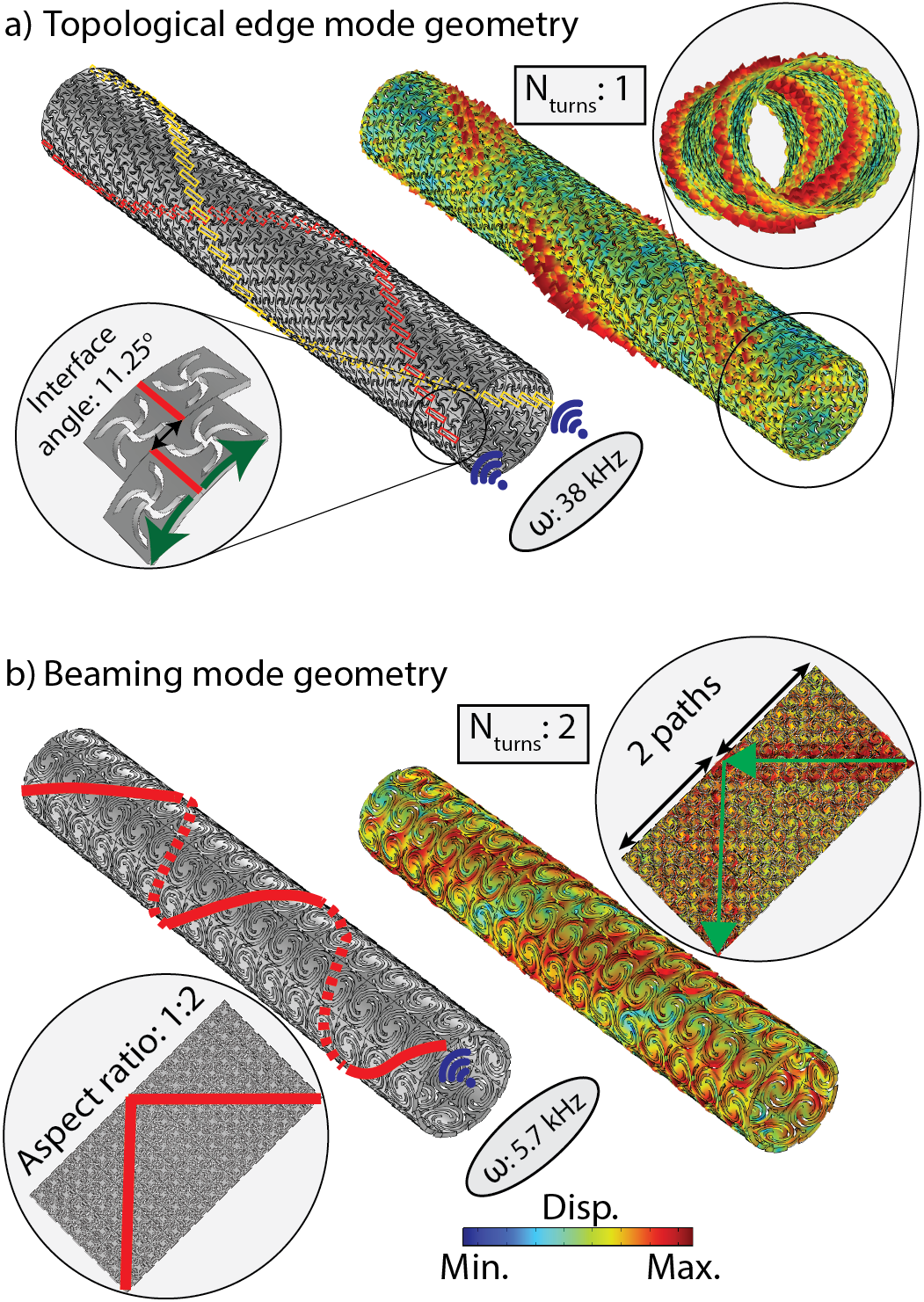}
\caption{\label{fig:Helical turns}\textbf{Helical waves:} a) Generation of a helical wave with 1 complete helical turn in curved cylinder of  Design 1 with topologically protected frequency at 38 kHz, by setting the interface angle between the rings to be $11.25^{o}$. The cylinder has two interface lines; each line makes a complete helical turn along the axial length of the cylinder. b) Generation of helical wave with 2 complete helical turns with Design 2 with a beaming frequency of 5.7 kHz, by setting the structure aspect ratio to be 1:2 (i.e., corresponding to a flat plate with 8x16 unit cells).}
\end{figure}

\begin{figure*}
\centering
\includegraphics[width= \textwidth]{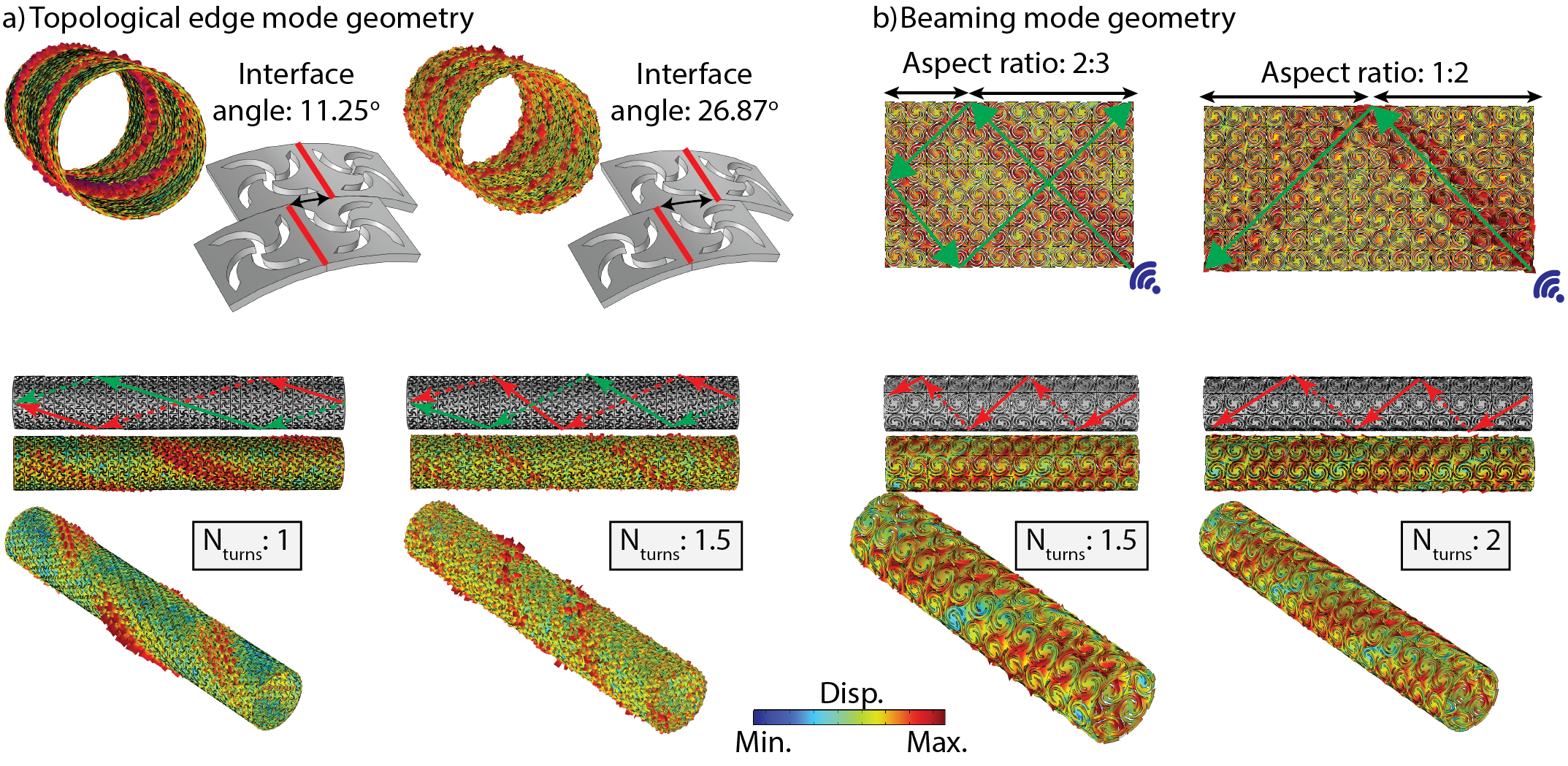}
\caption{\label{fig:5}\textbf{Tuning the number of helical turns:} a) Topologically protected edge mode with interface angle $11.25^{o}$ with one complete turn (left) changing into an interface angle $26.87^{o}$ making 1.5 helical turns (right). b) Tuning the number of helical turns of beaming mode (Design 2) by changing the aspect ratio of the corresponding flat plat. With an aspect ratio of 2:3, a helical wave of 1.5 complete turns is generated at the surface of the cylinder (left), while the geometry with aspect ratio of 1:2 gives results in a helical wave with 2 complete turns (right).}
\end{figure*}

\paragraph{Topological helix:} We start by constructing a cylinder from Design 1, where a topologically protected edge mode exits at 38 kHz [Fig. \ref{fig:Topological mode} c]. We create an interface on the surface of a cylinder by reflecting the spiral cut along the circumferential direction. This translate to having two versions of Design 1, left-handed and right-handed.  We repeat each design 8 times in a row over the circumferential direction to obtain two half rings. Both halves form a complete ring with two interface lines that are $180^{o}$ away from each other. We repeat the ring 32 times, while rotating each ring by an angle $\psi = 22.5^\circ$ in the axial direction to obtain a cylinder. The progressive rotating angle is  $iT\psi/2$, where $i = 1,...,32$ is the ring index in the axial direction, and $T = 1$ is the total number of helical turns per cylinder. We excite both interfaces from one side of the cylinder with a single frequency of 38 kHz. The propagating wave follow the helical interfaces along the surface of the cylinder, generating a helical wave at each path with one complete turn along the axial length of the cylinder [Fig. \ref{fig:Helical turns} a]. By changing the number of turns $T$ to 1.5, we generate one and half helix along the cylinder axis [Fig. \ref{fig:5}a].

\paragraph{Helical beaming:} We further expand the utilization of our framework using Design 2, this time with a beaming mode in a fully homogeneous cylinder [Fig. \ref{fig:Beaming mode} c]. The cylinder has 8 unit cells along the  circumference  and 16 axially corresponding to a flat structure with an aspect ratio (AR) of 1:2. This AR produces two paths on the diagonal of each half of the flat plate, with one reflection point on the side of the plate. Once curved, the flat path turns into two complete helical turns on the surface of the cylinder [Fig. \ref{fig:Helical turns} b]. We excite the cylinder from one side at 5.7 kHz revealing two complete turns of the wave propagating over the surface of the cylinder in the diagonal direction of each unit cell. Additionally, by changing the AR to 2:3, we obtain a helix with 1.5 turns [Fig. \ref{fig:5}b] .

\section{Discussion}

Our analysis shows the utility of spiral-based metamaterials to generate helical waves on the surface of a cylinder, and tune the number of turns either by controlling the interface angle in the case of topologically protected edge modes (Design 1), or controlling the aspect ratio in the case of beaming modes on a homogeneous (i.e., without interface) metamaterial cylinder (Design 2).

Within our platform, we study the dynamics of spiral-based metamaterials both in flat and curved settings. We consider various spiral designs that obtain different elastic band gap opening mechanisms. We explore the influence of curvature on the dynamical behavior of the metamaterials. We calculate the dispersion diagrams for both flat and curved unit cells and we observe a change in the band gap width and position as the unit cells transition from flat to curved cases. We validate the infinite unit cell dispersion diagrams with numerical and experimental frequency response functions (FRFs). We highlight two particular designs with interesting wave physics. Design 1 with topologically protected interface mode and Design 2 with strong wave beaming.  For Design 1, we observe a shift in  frequency, from flat to curved, of  the topologically protected edge mode as well as the beaming mode in Design 2. We demonstrate the progress of the shift in frequency for both modes, by plotting the mode shapes of the flat plates as they turn into fully curved cylinders. We harness the curvature of the metamaterials, to generate helical waves propagating along the surface of the metamaterial cylinders. We tune the number of helical turns in Design 1 cylinder  by changing the interface angle, and in Design 2 cylinder by changing the aspect ratio. Our findings demonstrate the possibility of helical wave generation with simple single point excitation, without the need to multiple polarizations or phase-locking facilitating less complexity in both design and excitation. Our framework could be beneficial for applications utilizing helical waves, such as non-destructive testing, crack detection, or energy harvesting.

\section{Methods}
\paragraph{Fabrication:} We fabricate the metamaterials using subtractive manufacturing (Full Spectrum PS24 Laser cutter). We use flat acrylic plastic sheets of thickness $2.4 mm$, and cylindrical tube with thickness $2.4 mm$  and outside diameter $55 mm$ . We excite the structures using piezoelectric bending disks (T216-A4NO-05), and we use a scanning laser Doppler vibrometer (Polytech 500-PSV) for the displacement measurements.
\paragraph{Dispersion calculations:} We solve the eigenvalue problem $(-\omega^2 \textbf{M} + \textbf{K}(\kappa))\bar{U} = 0$, where $\omega$ is the angular frequency, $\textbf{M},\: and \: \textbf{K}$ are the mass and stiffness matrices, respectively, $\kappa$ is the wavenumber and $\bar{U}$ is the eigenvector. Numerical simulations are conducted using finite element commercial package (COMSOL Multiphysics 6.2).
\paragraph{Frequency response functions (FRFs):} For flat structures, FRFs are calculated for a flat plate of 8x8 unit cells. We excite the plate at one corner with a chirp signal that sweeps over a specific frequency range, and we measure the transmitted displacement at the diagonally opposite corner. For the curved structures, we construct a metamaterial cylinders by repeating a curved unit cell 8 times circumferentially, producing a complete ring. Then we repeat this ring 8 times in the axial direction making a metamaterial cylinder. We excite the curved structure with a chirp signal at the bottom unit cell from one side, and measure the transmitted displacement at the top unit cell in the other axial side of the cylinder. The transmission is calculated as the output displacement divided by the input excitation displacement.

\begin{acknowledgments}
{We gratefully acknowledge the Air Force Research Laboratory, Materials and Manufacturing Directorate (AFRL/RXMS) for support via Contract/grant No. FA8650–21–C5711. Distribution A. Approved for public release AFRL-2025-1738: distribution unlimited.}
\end{acknowledgments}

\newpage
\beginsupplement
\newpage
\begin{widetext}
\newpage\hspace{-3mm}\Large{\textbf{Supporting Information: \\}}
\Large{\textbf{Harnessing curvature for helical wave excitation in spiral-based metamaterial structures}\\}

In this paper, we study the dynamics of spiral-based metamaterials, where changing spiral parameters can change the dynamical behavior of the structure [Fig. \ref{fig:S1}]. We consider four different designs where there exist different sources of band gap opening mechanism. For example, we consider spiral Designs 1 and 2, where the band gap opens due to Bragg scattering [Fig. \ref{fig:S1} a-b]. Additionally, we consider spiral Design 3, where the band gap opens due to localization of energy at a certain geometric feature (i.e., Local resonance) [Fig. \ref{fig:S1} c]. Finally, we consider spiral Design 4, where the band gap source appears to be the amplification of inertia [Fig. \ref{fig:S1} c] \cite{foehr2018spiral}. We consider flat and curved unit cells, where the unit cell size is $a = 21 mm$, density $\rho = 1185 kg/m^3$, Poisson's ratio $\nu = 0.37$, Young's modulus $E = 3 GPa$, and thickness $t = 2.4 mm$. For the curved unit cell, the mean radius of curvature $R_{\theta} = 26.5 mm$ (i.e., the radius of the cylindrical structures), and the number of unit cells in the circumferential direction is $N = 8$.
For Design 1, the spiral parameters are $R = 10 mm$, $r = 0$, $W = 2.4 mm$, and $\alpha = 45 ^{o}$, and $n = 0.25$. For Design 2, the spiral parameters are $R = 11.73 mm$, $r = 3.35$, $W = 1.4 mm$, and $\alpha = 45 ^{o}$, and $n = 0.65$.

\begin{figure}
\centering
\includegraphics[width= \columnwidth]{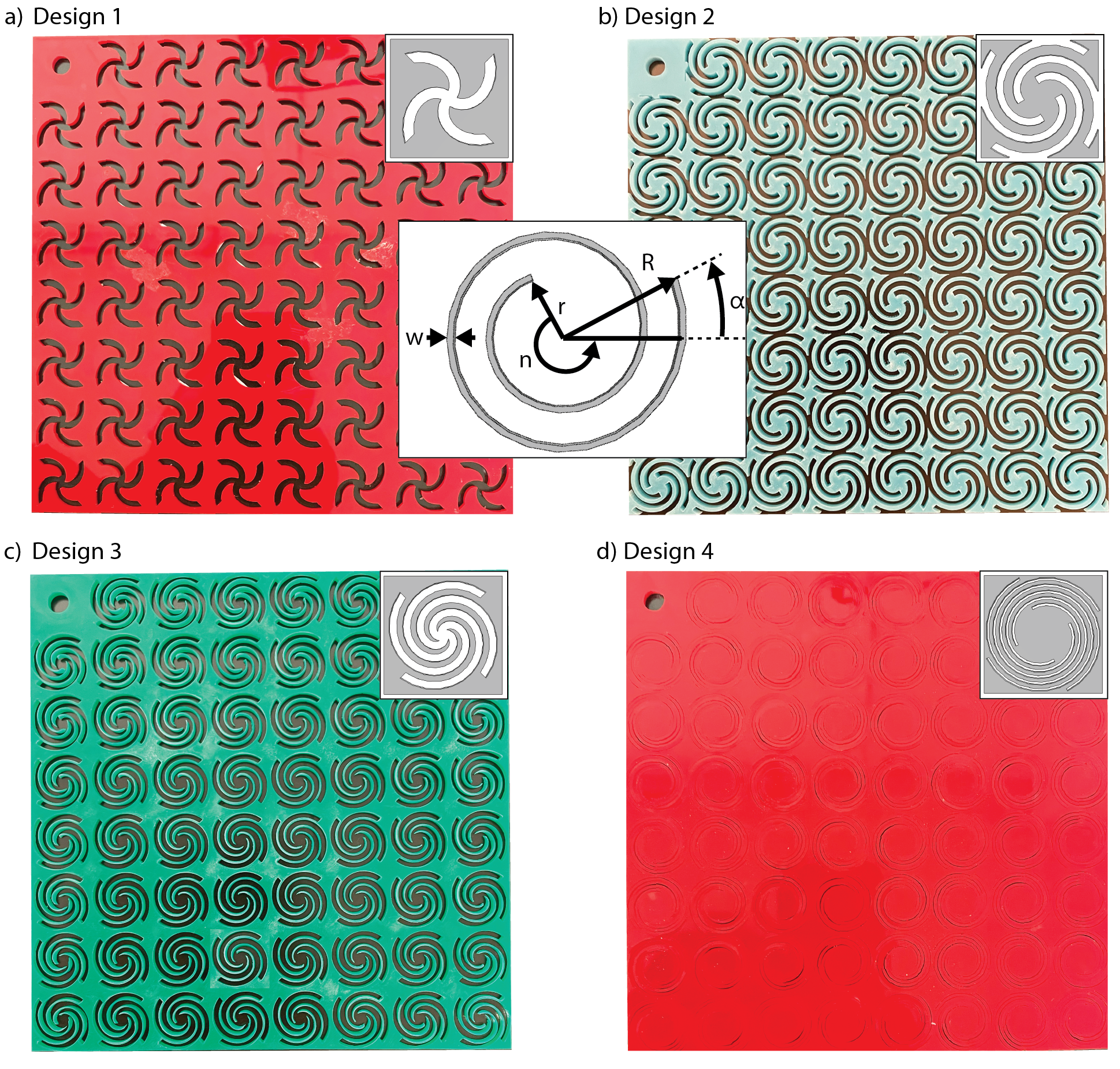}
\caption{\label{fig:S1}\textbf{Spiral geometries:} Spiral parameters including inner radius $r$, outer radius $R$, number of turns $n$, cutting width $W$, and orientation angle $\alpha$. a) Spiral Design 1 with the spiral parameters are $R = 10 mm$, $r = 0$, $W = 2.4 mm$, and $\alpha = 45 ^{o}$, and $n = 0.25$. b) Spiral Design 2 where the spiral parameters are $R = 11.73 mm$, $r = 3.35$, $W = 1.4 mm$, and $\alpha = 45 ^{o}$, and $n = 0.65$. c) Spiral Design 3 where the spiral parameters are $R = 10 mm$, $r = 0$, $W = 1.43 mm$, and $\alpha = 45 ^{o}$, and $n = 1$. and d) Spiral Design 4 where the spiral parameters are $R = 10.6 mm$, $r = 5 mm$, $W = 0.5 mm$, and $\alpha = 23 ^{o}$, and $n = 1.25$.}
\end{figure}

\begin{figure}
\centering
\includegraphics[width= \columnwidth]{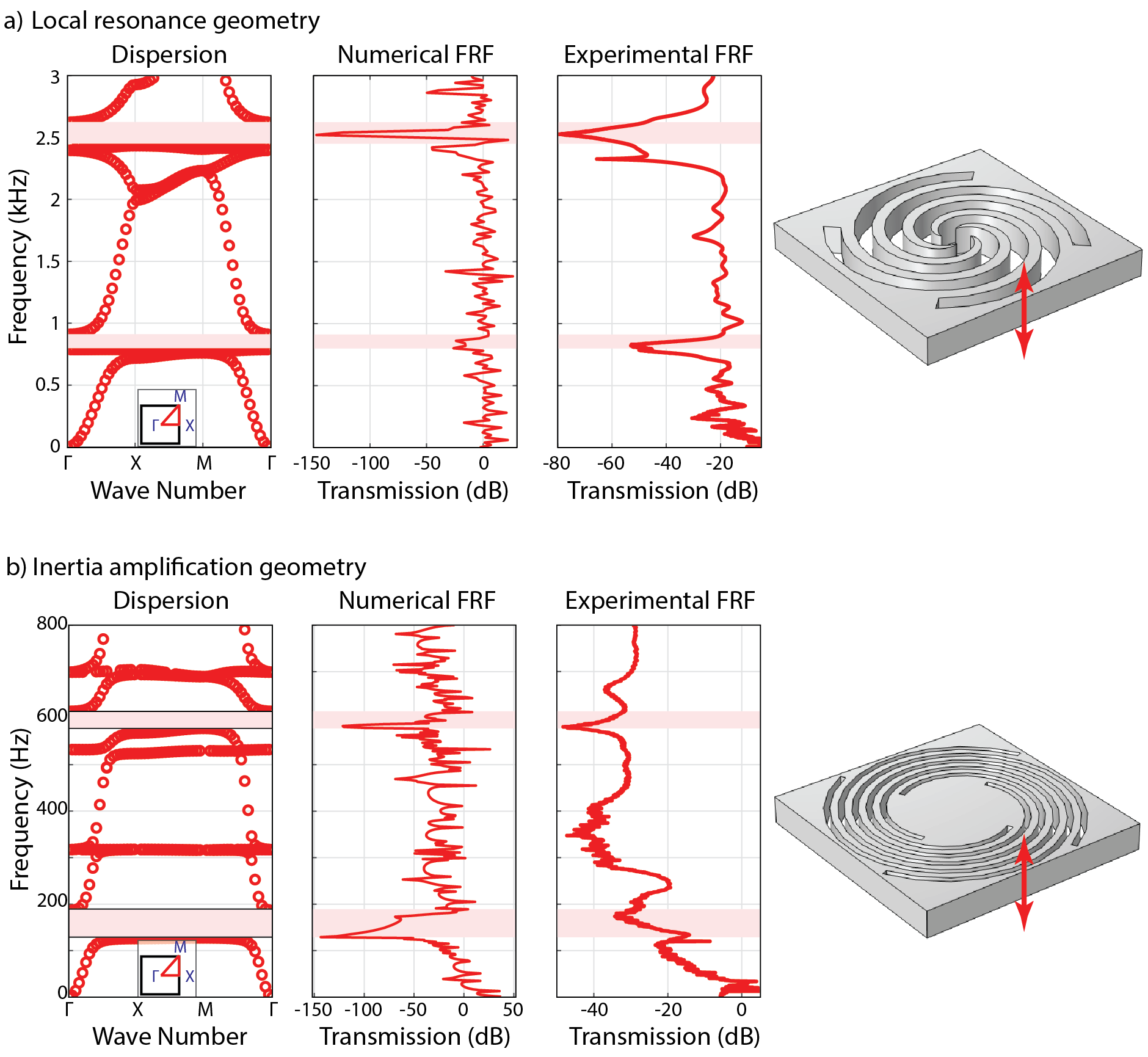}
\caption{\label{fig:S2}\textbf{Flat spiral designs 3 and 4:} Out-of-plane flat dispersion diagram, numerical frequency response function, and experimental frequency response function for a) spiral design 3 (Local resonance) in the frequency range 1 Hz to 3 kHz and b) spiral design 4 (Inertia amplification) in the frequency range 1 to 800 Hz.}
\end{figure}

\begin{figure}
\centering
\includegraphics[width= \columnwidth]{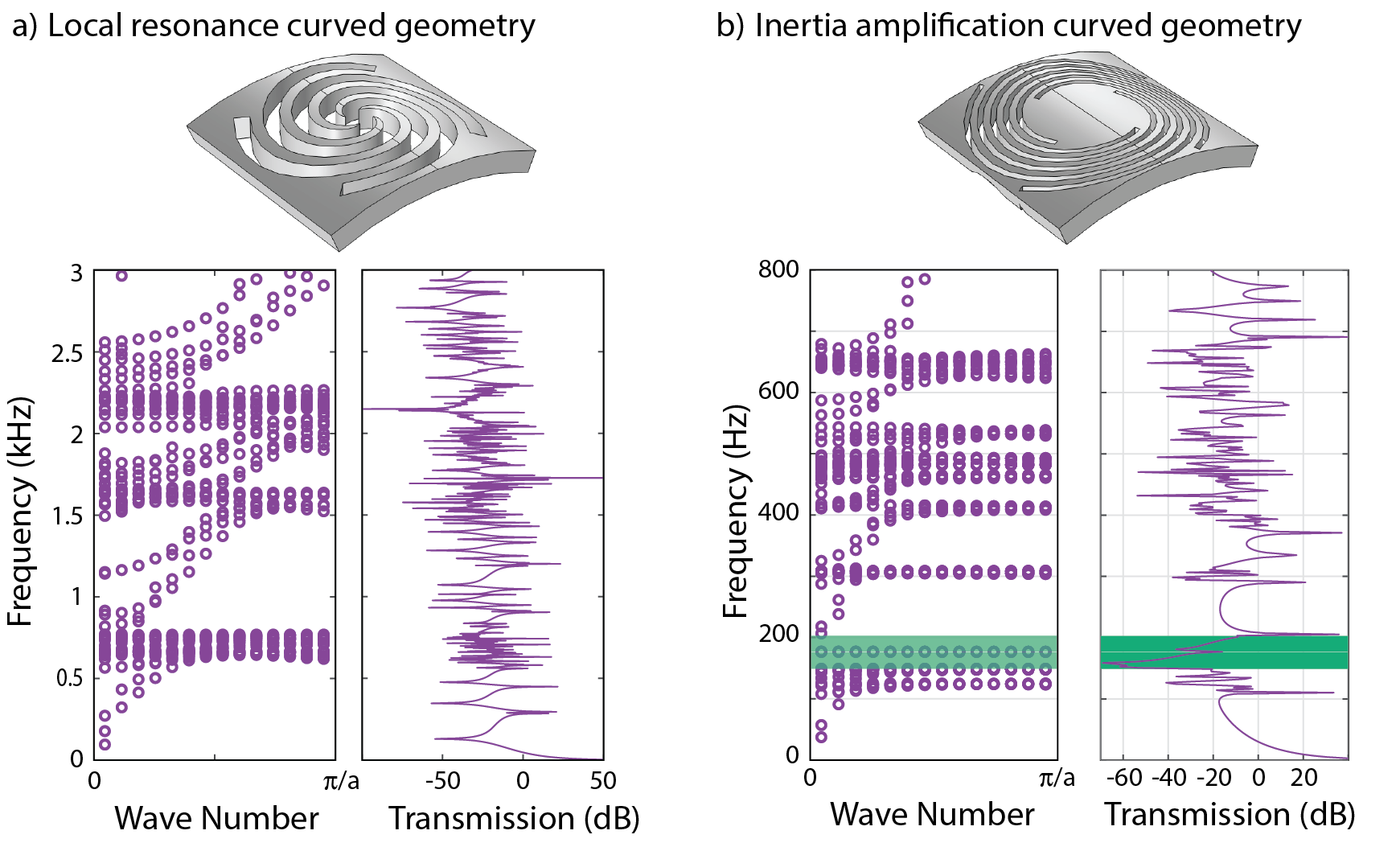}
\caption{\label{fig:S3}\textbf{Curved spiral designs 3 and 4:} Curved dispersion diagram and numerical frequency response function for a) spiral design 3 (Local resonance) and b) spiral design 4 (Inertia amplification). The unit cell of both designs is curved over an angle of $45^{o}$ making 8 unit cells in the circumferential direction. }
\end{figure}

\begin{figure}
\centering
\includegraphics[width= \columnwidth]{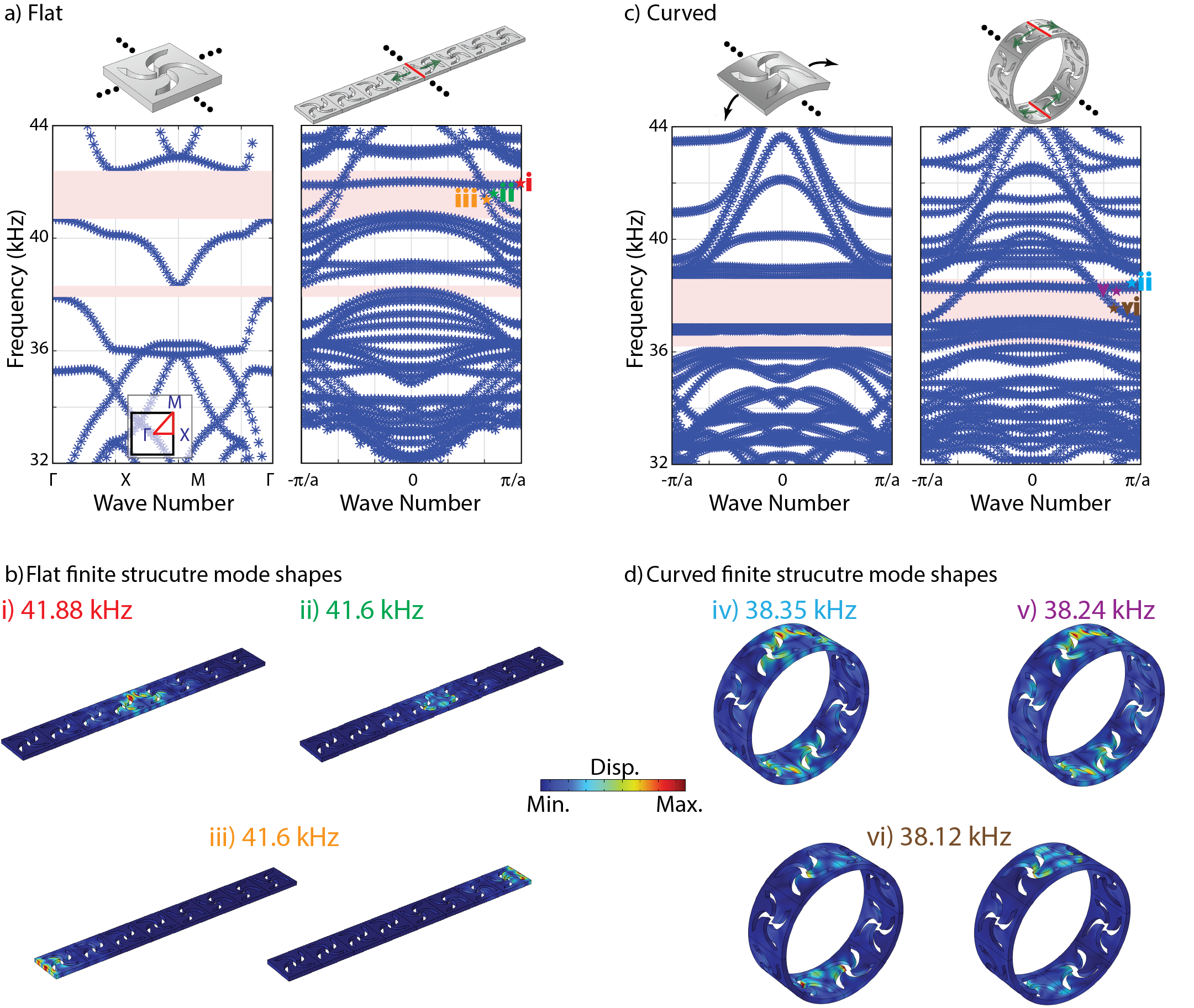}
\caption{\label{fig:S4}\textbf{Spiral design 1 topological edge mode:} a) Dispersion diagram for single flat unit cell periodic in two direction and for 1-D beam with interface line in the middle and periodic in one direction. b) Mode shapes at the beam dispersion bands inside the flat unit cell band gap showing localization of modes at the interface line and at the sides of the beam. c) Dispersion diagram for single curved unit cell periodic in the axial and circumferential directions and for a ring, with two interface lines, periodic in the axial direction. d) Mode shapes at the ring bands inside the curved unit cell band gap showing localization of the modes at the two interface lines.}
\end{figure}

\begin{figure}[!h]
\centering
\includegraphics[width= \columnwidth]{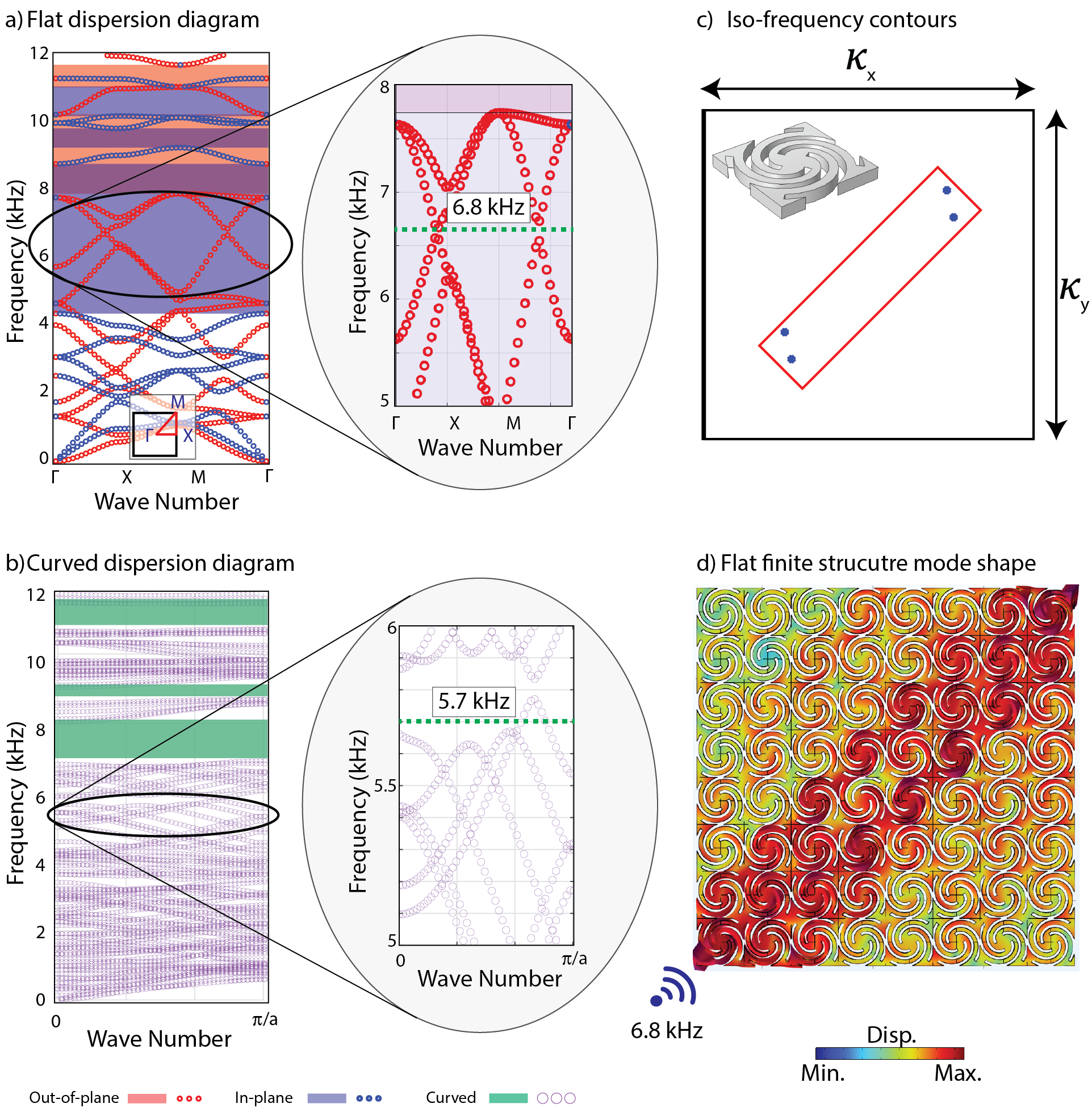}
\caption{\label{fig:S5}\textbf{Spiral design 2 beaming iso-frequency:} a) Dispersion diagram for flat unit cell of spiral design 2 in the frequency range 1 Hz to 12 kHz, with a zoomed-in view from 5 to 8 showing an intersection between 6.8 kHz and a beaming mode. b) Dispersion diagram for curved unit cell of spiral design 2 in the frequency range 1 Hz to 12 kHz with a zoomed-in view from 5 to 6 kHz showing a decrease in the frequency of the beaming mode due to curvature, to intersect with a frequency of 5.7 kHz. c) Flat unit cell Iso-frequency contours in reciprocal space showing projection of 6.8 kHz on a unit cell with a beaming mode taking place through the unit cell diagonal. d) Beaming mode shape a flat plate with 8x8 unit cells excite from the left bottom corner with 6.8 kHz, and the wave propagating through the diagonal of the plate.}
\end{figure}

For  Design 3, where the source of band gap opening is the the local resonance, the spiral parameters are $R = 10 mm$, $r = 0$, $W = 1.43 mm$, and $\alpha = 45 ^{o}$, and $n = 1$. We calculate the out-of-plane dispersion diagram for the flat unit cell, with an irreducible Brillouin zone (IBZ) of $\Gamma-X-M-\Gamma$. We observe two out-of-plane band gaps with central band gap frequencies $\omega_c$ 0.86 and 2.54 kHz, respectively, and band gap percentage ($BG \%$) of 13.1 and 6.8 \%, respectively. We validate the single-unit cell dispersion analysis by conducting numerical and experimental finite-structure analysis. We construct a flat plate by repeating 8x8 unit cells. We excite one corner of the plate in the out-of-plane direction, and measure the displacement at the diagonally opposite corner of the plate. We plot the FRF (displacement vs. frequency) both numerically and experimentally. The experimental FRF matches very well with the numerical one. Furthermore, both numerical and experimental FRFs are in very good agreement with the pass band and band gap regions predicted by the dispersion analysis [Fig. \ref{fig:S2} a].
We conduct a similar analysis for the flat spiral Design 4 (i.e., inertial amplification), where the spiral parameters are $R = 10.6 mm$, $r = 5 mm$, $W = 0.5 mm$, and $\alpha = 23 ^{o}$, and $n = 1.25$. We calculate the flat out-of-plane dispersion diagram with the irreducible Brillouin zone (IBZ) $\Gamma-X-M-\Gamma$. We observe two out-of-plane band gaps with $\omega_c$ of 160 and 596 Hz, respectively, and $BG \%$ of 38 and 6.2 \%, respectively. In addition, we plot the numerical and experimental FRFs for a 8x8 flat plate. Both numerical and experimental FRFs show very good matching with each other, and very good agreement with band gaps from the dispersion diagram, which validates the single unit cell dispersion analysis both numerically and experimentally [Fig. \ref{fig:S2} b].

In order to investigate the effect of curvature on the dynamical behavior of Designs 3 and 4, we calculate the dispersion diagrams for the same unit cells, but in curved settings. We start with Design 3, by curving the unit cell over a curvature angle of $45^{o}$. We calculate the dispersion diagram in the axial direction for all circumferential modes (m = 0 to 4) in the frequency range up to 3 kHz. The dispersion diagram shows that both band gaps in the case of the flat unit cell completely disappear, making the considered dispersion range complete transmission. We validate the dispersion analysis with finite FRF for a cylinder with 8 unit cells around the circumference and 8 unit cells in the axial directions. The numerical FRF shows a complete transmission region without attenuation in the considered frequency range [Fig. \ref{fig:S3} a].
For Design 4, we calculate the dispersion diagram for the curved unit cell in the axial direction for all circumferential modes (m = 0 to 4) in the frequency range up to 800 Hz. We observe two band gaps with $\omega_c$ of 162.5 and 190 Hz, and $BG \%$ of 17 and 14 \%, respectively. We validate the curved single unit cell dispersion with numerical finite-structure FRF. The numerical FRF shows good agreement with the predicted band gaps from the dispersion diagram [Fig. \ref{fig:S3} b].

In order to validate the existence of a topologically protected edge mode in design 1, we calculate the dispersion diagram for a quasi-infinite structures for both flat and curved cases. We start with the flat case, where we construct a 1-D  beam with 8 unit cells of Design 1, with a reflection line in the middle of the beam, making 4 unit cell right-sided spirals, and the other 4 unit cell left-sided spirals. The beam has infinite periodicity in the axial direction. We calculate the dispersion diagram for a wave vector from $-\pi/a$ to $\pi/a$ in the direction of the periodicity of the beam. The dispersion diagram of the beam shows the existence of two degenerate trivial modes completely inside the band gap, in addition to other degenerate modes crossing the dispersion mode bulk below the band gap to the mode bulk above the gap [Fig. \ref{fig:S4} a]. We plot the mode shapes at the trivial degenerate modes and the degenerate modes crossing the band gap. The mode shapes show localization of the mode at the interface or at the edges of the beam [Fig. \ref{fig:S4} b]. For the curved case, we curve the beam considered in the flat case to construct a ring with two interface lines. We calculate the dispersion diagram of the ring with interface lines considering periodicity in the axial direction over a wave vector $-\pi/a$ to $\pi/a$. The dispersion diagram shows two degenerate trivial modes inside the band gap and two degenerate modes crossing the band gap [Fig. \ref{fig:S4} c]. The mode shapes at the bands inside the band gap and the bands crossing the band gap show localization of mode at the two interface lines in the ring [Fig. \ref{fig:S4} d]. For both cases flat and curved, the dispersion diagram of the quasi-finite structures and mode shapes confirm the existence of topologically protected edge mode.

In order to validate the existence of a beaming mode at 6.8 kHz for Design 2, we calculate the iso-frequency contours in the full reciprocal space of the flat unit cell. The zoomed-in view of the dispersion diagram in the frequency range of 5 to 8 kHz for the flat unit cell shows an out-of-plane mode that intersects with the frequency of 6.8 kHz at two points between $\Gamma$ and $X$ in the reciprocal space [Fig. \ref{fig:S5} a]. The iso-frequency contours at the frequency of 6.8 kHz shows the projection of the frequency 6.8 kHz on a single unit cell, in the reciprocal space, which shows the existence of the eigen solutions at 6.8 kHz along the unit cell diagonal [Fig. \ref{fig:S5} b]. For the case of the curved unit cell, the zoomed-in view of the curved dispersion diagram shows a mode that intersects with the frequency of 5.7 kHz at two points of the wave vector in the axial direction [Fig. \ref{fig:S5} c]. We plot the mode shape of the flat plate with 8x8 unit cells, excited from one corner at 6.8 kHz, and the wave only propagates through the diagonal of the plate towards the diagonally opposite corner [Fig. \ref{fig:S5} d].\\

\end{widetext}
\end{document}